# Non-Parametric Dynamical Analysis of Globular Clusters: M15, 47 Tuc, NGC 362, and NGC 3201

Karl Gebhardt

Department of Physics and Astronomy, Rutgers, The State University, Box 0849
Piscataway, NJ 08855-0849

gebhardt@physics.rutgers.edu

and

Philippe Fischer

AT&T Bell Laboratories, 1d-316, 600 Mountain Ave, Murray Hill, NJ 07974

philf@physics.att.com

## ABSTRACT

We use radial velocities of member stars and cluster surface brightness profiles to non-parametrically determine the mass density profiles and isotropic phase-space distribution functions $f(E)$ for the globular clusters M15 (NGC 7078), 47 Tuc (NGC 104), NGC 362, and NGC 3201. Assuming isotropy and using the velocity dispersion and surface brightness profiles, the Jeans equation uniquely determines the mass density profile. For M15 and 47 Tuc, the slopes of the mass density profiles beyond $2.0'$ are similar to the theoretical predictions for core-collapse, whereas the density profiles for NGC 362 and NGC 3201 are steeper than the predicted slope. In the two centrally-concentrated clusters, M15 and 47 Tuc, we find that the mass-to-light ratios (M/L's) reach minima around $1'$, and increase by more than a factor of four towards the cluster centers. For the two less centrally concentrated clusters, the M/L decreases monotonically all the way into the center. All four clusters exhibit an increase in the M/L's in their outer parts. If the variations in the M/L's are due to equipartition of energy between different mass stars, then we attribute the central increases to massive remnants and the outer increases to low-mass stars (m< $0.3 M_\odot$). By applying the crude approximation of local thermodynamic equilibrium, we derive the present-day mass function for each cluster. In the central 2–3 parsecs, 0.7–1.5 $M_\odot$ objects provide the bulk of the cluster mass. These results may be sensitive to the assumption of isotropy in the stellar velocities.

We derive phase-space distribution functions $f(E)$ for the clusters and find significant disagreement with the best-fit King distribution functions. For NGC 362 and NGC 3201, there are significantly more low $E$ objects than predicted by King models. The $f(E)$ for M15 and 47 Tuc are similar to each other but show differences from the King models. The techniques described in this paper can be used for other dynamical systems, such as galactic nuclei and clusters of galaxies.

*Subject headings:* Globular Clusters, Stellar Systems (Kinematics, Dynamics)

## 1. Introduction

The conventional approach used to constrain the dynamics of globular clusters involves comparison of the clusters' surface brightness or surface density profiles with single and multi-mass Michie-King (MK) models (Michie 1963, King 1966, Da Costa & Freeman 1976, Gunn & Griffin 1979). Radial velocities are usually only used to determine the cluster masses as the datasets have traditionally been too sparse to offer further dynamical constraints. The MK approach has the advantage of providing reasonable results for noisy, sparsely sampled data, and avoids the need to deproject the data, which results in noise amplification (King 1981). The disadvantage is that the results can be biased by the functional form of the assumed models; dynamically important deviations from the model may be missed. Recently, with the development of absorption-line Fabry-Perot and high-resolution multi-fiber spectroscopy, the radial velocity datasets have improved in both quantity and quality, making it possible and desirable to use non-parametric techniques to derive meaningful dynamical information from the kinematic as well as the morphological data.

In this paper, our goal is to infer the form of the cluster gravitational potential $\Phi(r)$ and the isotropic stellar distribution function $f(E)$ given observations of the surface densities and radial velocity dispersion profiles (VDPs) of a "tracer" population. Merritt (1993a, b) has pointed out that this problem can be solved non-parametrically via a regularized algorithm. Our approach is similar, except that we will impose smoothness in a rather different, though equally non-parametric, way. We will assume isotropic, non-rotating orbital distribution functions (DFs); a future paper will discuss anisotropic DFs.

In Sec. 2 we give a step-by-step description of the non-parametric techniques and in Sec. 3 we describe some simulation-based tests. Sec. 4 applies our technique to four cluster datasets and derives density profiles. Sec. 5 derives the phase-space distribution functions for the four clusters and compares them to King models. In Sec. 6 we estimate the present-day mass functions. In Sec. 7 we summarize and discuss our results.

## 2. Determining the Mass Density

In order to obtain mass density profiles, we require cluster surface brightness profiles (SBPs) (or surface density profiles) and stellar radial velocities for a tracer population of the cluster. For globular clusters, the giant and turn-off stars provide both the SBPs and the radial velocities. We estimate a smooth velocity dispersion profile (VDP) from the radial velocities using the LOWESS fitting procedure (Cleveland & McGill 1984, Gebhardt et al. 1994a). The SBP and VDP are then deprojected using the Abel integrals:

$$\nu(r) = -\frac{1}{\pi} \int_r^{R_{max}} \frac{d\,I(R)}{dR} \frac{dR}{\sqrt{R^2 - r^2}} \quad \text{and,} \tag{1}$$

$$\nu(r)v_r^2(r) = -\frac{1}{\pi} \int_r^{R_{max}} \frac{d\,[I(R)\sigma_p^2(R)]}{dR} \frac{dR}{\sqrt{R^2 - r^2}} \quad , \tag{2}$$

where $\nu(r)$, $I(R)$, $\sigma_p(R)$, and $v_r(r)$ are, respectively, the luminosity density, SBP, and projected and deprojected VDPs of the tracer population. The integral extends out to the tidal radius, but, since it is not usually possible to obtain measurements out to such large radii, we extrapolate the integrands from the last radius where both the SBP and VDP are known. This will not significantly impact the derived mass density in the inner regions in which we are most interested, since for globular clusters $I(R)$ decreases steeply in the outer parts and these regions will contribute little ($< 10^{-10}$) to the integrals. For other dynamical systems where the fall-off is less steep, such as clusters of galaxies, more care would have to be taken in the integral evaluations.

Once we have the deprojected profiles for the tracer population, we can use the isotropic non-rotating Jeans equation to estimate the total mass profile, $M(r)$, and total mass density $\rho(r)$,

$$M(r) = -\frac{rv_r^2}{G} \left( \frac{d\,\ln\nu}{d\,\ln r} + \frac{d\,\ln v_r^2}{d\,\ln r} \right) \tag{3}$$

$$\rho(r) = \frac{1}{4\pi r^2} \frac{dM}{dr} \,. \tag{4}$$

Equations 1 through 4 employ a total of two and a half derivatives of the SBP and VDP, both of which are intrinsically noisy. Since the derivatives will be even noisier, a consistent way of evaluating these quantities is to carry out the mathematical operations defined by equations (3) and (4) directly on the smooth estimates of $\nu$ and $v_r^2$ (e.g. Wahba 1990, p.19). Smoothing is achieved using the GCVSPL program (obtained via netlib; send the message "send index" to netlib@research.att.com), a spline smoother where the smoothing parameter is chosen by generalized cross validation (GCV). We also use GCVSPL to obtain an analytical functional form for the integrands which allows for more accurate numerical integration. The smoothing is carried out in log space to lessen the dynamic range in the functions which GCVSPL will estimate. The choice of the smoothing parameter can have a significant effect on the results; GCV will not provide the proper degree of smoothing unless the measurement uncertainties are accurately determined. Also, proper smoothing for the data is not proper smoothing for derivatives (Scott 1992, p.131). For situations where the uncertainties are not reliably estimated we choose the smoothing which gives the smoothest estimate of the underlying distribution without introducing a significant bias. The bias can be determined by using simulated data (Sec. 3) or through eye estimates.

Figs. 1a-d show the projected and deprojected luminosity and velocity dispersion profiles for 47 Tuc. The SBP comes from Meylan (1988), and the VDP from Gebhardt et al. (1994b). The VDP is determined from the LOWESS technique, as noted above, and the details of the technique are given in Gebhardt et al. (1994b).

Smoothing can result in biased estimates of the radial profiles. This can be limited by reducing the smoothing parameter in the splines, which may result in unacceptably noisy output. Alternatively, one can use heavy smoothing and correct for the bias by using a bootstrap procedure which will be discussed below. The justification for using heavy smoothing is that we believe the underlying mass density and luminosity density distributions should be a smooth function of radius, at least on scales of tenths of parsecs. In Fig. 2, we plot the estimated mass density and mass-to-light (M/L) profiles for 47 Tuc with and without bias-correction. The M/L is obtained by dividing the mass density, plotted in Fig. 2, by the luminosity density in Fig. 1b. The projected VDP and SBP continue to both smaller and larger radii than are plotted for the mass density, but we only plot the mass density over ranges of radii for which it is reasonably constrained.

The bias and confidence bands were determined using 1000 bootstrap re-samplings. At the position of each star with a measured radial velocity, we generated an artificial radial velocity drawn randomly from a Gaussian distribution with a standard deviation equal to the projected model velocity dispersion at that point, and we then add the measurement uncertainty. The original estimate of the SBP is used for each realization since the dominant source of uncertainty in the mass estimation comes from the VDP for the clusters presented here. We then determined the mass density profile for each realization as outlined above. From the ensemble of realizations we inferred the mode and the 90% confidence band for the simulated mass density profile. Since we know the true profile of the simulated data, we can estimate the smoothing bias and correct the mass distribution. The confidence bands require double the correction since the simulations have a bias from the technique and a bias from the original estimate (cf. Scott 1992, p.259). The difference between the bias-corrected and original estimate is generally small except near the edges of the data. We have assumed that the velocity distribution at each radius is a Gaussian, which is probably not the case due to the tidal cutoff imposed by the Galaxy. A fully non-parametric technique for estimating the confidence bands would need to include a proper estimate of the velocity distribution at each radius, which is presently unknown. However, effects from the tidal cutoff are not large in our main region of interest, near the cluster center.

Here is a summary of the steps used to obtain the mass density profile:

1. Calculate smoothed estimates of the SBP and its derivative – uses GCVSPL.
2. Determine projected VDP, $\sigma_p^2(R)$ from radial velocities using LOWESS.
3. Calculate continuous derivative of $I(R)\sigma_p^2(R)$ – uses GCVSPL.
4. Deproject $I(R)$ and $\sigma_p^2(R)$ using Abel integrals (Eqns. 1 and 2) to yield the luminosity density, $\nu(r)$, and the VDP, $v_r^2(r)$.
5. Calculate continuous derivatives for $\ln[\nu(r)]$ and $\ln[v_r^2(r)]$.
6. Jeans' equation (Eqn. 3) gives the mass profile.
7. Eqn. 4 yields the mass density.
8. The M/L profile is straight-forwardly calculated by dividing the mass density by the luminosity density.

The steps above consist of one call to the smoothing routine LOWESS and two smoothing calls to GCVSPL. The smoothing which occurs in the LOWESS routine is by far the most likely to yield a bias. We generally use a smoothing parameter of 0.5–0.6 in the LOWESS estimate, with smaller samples requiring a larger smoothing parameter. The smoothing parameter is the fraction of data points which are used to estimate the value of the profile at a particular projected radius (see Gebhardt 1994a).

## 3. Simulations

In order to test our technique we used simulated data drawn from multi-mass Michie-King (MK) models with power-law mass function exponents ranging from $0 \leq x \leq 2$ (Salpeter – $x = 1.35$), and anisotropy radii in the range $3 \leq r_a/r_s \leq \infty$, where $r_s$ is the scale radius (see Fischer et al. 1993 for a description of the models used). For each MK model we generated three-dimensional positions and space velocities which were projected and measurement uncertainties added to yield radial velocities. Simulations were carried out with either 500 or 1000 radial velocities, and each was analyzed as described above to produce mass density profiles which could then be compared to the known profiles. We assumed that the SBP was error-free since the uncertainties in the velocity dispersion profile tend to dominate. Since there were no significant differences between the 1000 and 500 radial velocity sets, aside from larger confidence bands on the latter, we will discuss only the former.

Fig. 3 compares the input mass density and M/L profiles with the average of 20 derived simulations with $x = 0$ and $r_a = \infty$. The two pairs of profiles are in excellent agreement and the actual profiles are well within the 90% confidence bands of the mass density and M/L estimates. Fig. 4 plots the results for one realization with $x = 2$ and demonstrates the amount of uncertainty with a sample of 1000 points. In Fig. 5 we use the same concentration and mass function slope as in Fig. 4, and show the effect that anisotropy will have on the derived profile, where we have assumed isotropy.

To summarize the results of the simulations, the isotropic model simulations always resulted in excellent agreement between the derived and known mass density profiles. The anisotropic models, however, all exhibited a common systematic feature: the derived density profiles tended to increase towards the cluster center more rapidly than the true profiles. This is not surprising since an anisotropic DF and a central mass cusp have similar effects on the projected VDP; both result in a steep increase in the projected VDP at the cluster center. Based on our simulations, the observed increase in the M/L is usually insignificant given the size of the confidence bands, which are larger for the anisotropic MK models than for the isotropic models. Another feature that the anisotropic models exhibit is that the mass density appears to be underestimated in the outer parts of the cluster, although once again, this is usually not significant. In the intermediate regions, the mass density and M/L profiles appear to be well-estimated even for small values of the anisotropy radius. This is surprising since we are assuming an isotropic DF for the Jeans' equation. Based on the simulations we conclude that the non-parametric modeling provides accurate mass density estimates.

## 4. Globular Cluster Data

We have non-parametrically modeled four globular clusters: M15, 47 Tuc, NGC 362, and NGC 3201. For M15, the 253 radial velocities come from Peterson et al. (1989) and Gebhardt et al. (1994a), and the SBP is from Grabhorn et al. (1992). For 47 Tuc, the 640 radial velocities are from Mayor et al. (1983), Meylan et al. (1991), and Gebhardt et al. (1994b), and the SBP is from Meylan et al. (1988). For NGC 362, the 201 radial velocities and the SBP are from Fischer et al. (1993). We will also discuss results from NGC 3201, the details of which are given in Côté et al. (1993, 1994).

The estimated mass density profiles of the four clusters are shown in Fig. 6. Fig. 7 shows the estimates for the M/L's. The density and M/L profiles for NGC 3201 are the same as given in Fig. 8 of Côté et al. (1994). For M15, Fig. 6 shows a region from $0.5'$ to $1.0'$ in which the lower confidence band of the density estimate becomes very uncertain. This is due to the sharp increase in the velocity dispersion profile which was found both by Peterson et al. (1989) and Gebhardt et al. (1994a). This large uncertainty reflects the need to obtain large amounts of data in the inner regions to adequately constrain the mass profile. For comparison, we have plotted the theoretically-predicted slope for core-collapse clusters, $d\ln\rho/d\ln r = -2.23$ (Cohn 1980), which will be discussed in Sec. 7.

The M/L profiles of the two centrally-concentrated clusters, M15 and 47 Tuc, are remarkably different from those of NGC 362 and NGC 3201. While the latter two have monotonically decreasing M/L towards the center, the first two have minima approximately $1'$ from the center and increase inwards towards the centers and outwards towards the tidal radii. The implications of this will be discussed in Sec. 6.

## 5. Phase-Space Distribution Functions

With the mass and luminosity density estimates we can calculate the phase-space distribution function for the tracer population. The potential is obtained by solving Poisson's equation,

$$\Phi(r) = -4\pi G \left[ \frac{1}{r} \int_0^r \rho(r') r'^2 dr' + \int_r^\infty \rho(r') r' dr' \right] \tag{5}$$

Eddington's equation then yields the phase-space distribution function, $f(E)$,

$$f(E) = \frac{1}{\sqrt{8}\pi^2} \frac{d}{dE} \int_E^0 \frac{d\nu}{d\Phi} \frac{d\Phi}{\sqrt{\Phi - E}}. \tag{6}$$

We do not know $\rho(r)$ all of the way into the center, but if we assume a constant $\rho$ inside of our last measured value then the contribution to $\Phi$ from this region is less than 1% of the total, so we only introduce a small error by assuming a constant $\rho$ in the inner region.

The resulting $f(E)$ and the 90% confidence bands for the four clusters are plotted in Fig. 8 along with the best-fit isotropic, multi-mass MK model (dashed line). The parameters used for the MK models are given in Table 1, where we give the central density (col.2), scale radius (col.3), tidal radius (col.4), total mass (col.5), and mean stellar mass at the center (col.6). For each cluster, the best isotropic MK models fit as well as or better than the best anisotropic MK models (Fischer et al. 1993, Côté et al. 1994, Pryor 1994). To test our $f(E)$ determinations we performed the same calculations on the simulated data sets described in Sec. 3 and found consistency between the known and derived values.

## 6. Mass Functions

The variation of the M/L profiles (Fig. 7) suggests a change in the stellar populations as a function of radius. We see an increase in the M/L in the central 1.0 parsec for the two centrally-concentrated clusters, and for all four clusters there is an increase in the outer regions. In globular clusters there are two types of unseen matter which will tend to raise the M/L: stellar remnants (i.e. white dwarfs, neutron stars, and black holes), and low-mass (m< 0.3 $M_\odot$) stars. Energy equipartition should result in the migration of heavy remnants to the central regions and low-mass stars to the outer parts of the clusters, consistent with our M/L profiles. It is desirable to quantify these results.

In order to determine the mass functions we require the mass density profile for each mass group, which we can estimate since the sum of the individual mass densities must equal the total derived mass density. The shape of the individual mass density profiles can be obtained through the Jeans equation provided the VDP is known for that particular mass group. Thus, a relation between the VDP for the different mass groups will lead to an estimate of the mass function. The best way to do this is by using realistic multi-mass evolutionary models (i.e. N-body or Fokker-Planck models). Such modeling is beyond the scope of this paper. Instead we will use the *very crude* approximation of local thermodynamic equilibrium (LTE). LTE implies a simple scaling relationship between the known tracer velocity dispersion and the dispersion of objects with different masses. The LTE approximation is probably not strictly valid (Merritt 1981, Inagaki & Saslaw 1985), but is quite reasonable in the cluster centers where the potential is the deepest.

Given our assumption of LTE, the velocity dispersion profile can be calculated for any mass. We use the VDP measured from the individual stellar velocities, which are all giants, and scale according to the mass ratio (i.e. $m\,v^2(r) = m_g v_g^2(r)$). We have taken into account the cluster escape velocities which affect the lowest mass stars. The giants are assumed to have mass 0.7 $M_\odot$. The Jeans equation gives the shape of the number density profile for each mass class $n_m(r)$:

$$\frac{1}{n_m}\frac{dn_m}{dr} = -\frac{m}{m_g v_g^2}\frac{d\Phi}{dr} - \frac{1}{v_g^2}\frac{dv_g^2}{dr}. \tag{7}$$

There is a normalization factor, $a_m$, for each $n_m$ which is not given by equation 7 and must be determined by using the cluster mass density profiles; the sum of the individual mass densities must equal the total mass density. This is done by minimizing the quantity

$$\sum_i \left\{\log\left[\rho(r_i)\right] - \log\left[\sum_m a_m m\, n_m(r_i)\right]\right\}^2 + \lambda P(a), \tag{8}$$

where $P(a)$ is a penalty function, and $\lambda$ is the smoothing parameter. The penalty function is necessary to overcome the degeneracy in assigning the relative normalizations for objects of similar mass. Without it, a simple minimization will lead to artificial noise in the mass function. The penalty function is a measure of the smoothness of the mass function. A standard form for the penalty function is the square of the second derivative (Silverman 1986, p.117):

$$P(a) = \int_m \left(\frac{d^2 \log a}{d \log m^2}\right)^2 d\log m, \tag{9}$$

where the second derivative is evaluated numerically.

The parameter $\lambda$ is used as the relative weight between the fit to the mass density and the smoothness of the mass function. Generally, one tries to provide as much smoothing as possible to the mass function without significantly degrading the fit to the mass density profile.

The number density profiles are integrated to generate the total numbers of objects per solar mass in each mass class (Fig. 9). The mass classes are shown as the solid circles which lie on the solid line for each cluster. We only evaluate the mass functions at each of the mass classes and we have connected the dots to aid in representation. We have divided the clusters into three radial regions: the first contains the inner 25% of the mass, the second is the next 25% of the mass, and the last contains the 50–70% mass fraction. These mass fractions are not fractions of the total cluster mass but instead are fractions of the total mass in the regions for which we have density estimates. The mass density profiles for M15 and 47 Tuc do not extend much beyond the half-mass radius.

The first point to notice is that the central regions contain a significantly higher fraction of heavier objects than the outer regions. This is not solely due to our assumption of LTE. LTE will cause the heavier objects to have more steeply declining profiles, but the ratio of the heavy to light objects is due to the normalization to the density profile. The sharp increase in the density profile seen in M15 and 47 Tuc at about 1.0′ is what drives the need for high-mass objects, since their number density profiles have the steepest decline.

Richer et al. (1990) have used deep star counts to determine mass functions for M13, M71, and NGC 6397. Their derived mass functions extend from 0.8–0.1 $M_\odot$. They found evidence for steep increases in the mass functions for masses less than 0.4 $M_\odot$ at large projected radius, and estimated power-law mass function exponents in the range of 0.5–2.7. Hesser et al. (1987) derive a luminosity function for 47 Tuc that is best fit with an exponent of 0.2. Although our results are not directly comparable since we do not extend to such large radii, the slopes we derive are similar, but slightly higher, with a range from 1.3–3.0. We do stress, however, that the low-mass end of our derived mass functions are the most uncertain. The low-mass stars deviate the most from LTE (Inagaki & Saslaw 1985) and their VDPs are affected the most by the truncation due to the cluster escape velocity. We also see an increase in the mass function towards the heavier masses, most likely due to remnants, which do not show up in the luminosity function studies (yet!), and make comparisons with the luminosity functions difficult. The mass functions imply a large number of stellar remnants ($0.7 \leq M (M_\odot) \leq 1.5$), consistent with fairly shallow initial mass functions.

Using measurements of pulsar accelerations for M15, Phinney (1992, 1993) found a central M/L of about 2.5 $M_\odot/L_\odot$ and a central mass density of $> 2 \times 10^6$ $M_\odot/pc^3$. The central M/L is similar to our innermost value at 0.2′ but our innermost mass density value is 100 times lower. Phinney's measurement is of the central density and it is not surprising that he finds a much higher value than we do, given the small core radius for M15. By comparing his pulsar observations with Fokker-Planck models, Phinney estimated that objects of about 1 $M_\odot$ dominate the central regions, consistent with our results.

We stress that the mass function estimates are calculated using the assumption of LTE. We have also assumed isotropic velocity dispersions. Therefore, the mass functions we derive should only be used for order of magnitude estimates. The details of the mass functions at larger radii (the dotted lines in Fig. 9) are the most suspect since that is the region which will have the strongest deviations from LTE. We do note, however, that we see a very significant peak at and above the giant mass (0.7–1.5 $M_\odot$), which is quite robust since the high-mass stars will be closest to LTE. Ideally, a better model for the temperature dependence of the mass classes should be used instead of the simple assumption of LTE.

## 7. Conclusions

In this paper we have presented a new non-parametric technique for analyzing the dynamics of globular clusters. This technique is free of the biases inherent in parametric model-fitting. We have applied this technique to four globular clusters, M15, 47 Tuc, NGC 362 and NGC 3201. We have assumed non-rotating velocity distributions. Future work should incorporate rotation for a complete dynamical analysis. For the four clusters studied, the rotation is small compared to the dispersion and will have a small effect on the present dynamical analysis.

We have compared the non-parametric mass densities with the theoretical slopes for core-collapse clusters, $d\log(\rho)/d\log(r) = -2.23$, for radii outside of the core radius (Cohn 1980). The radial extent for the constant slope of the mass density is dependent on the age of the cluster, and can extend over several decades of radius during late epochs of core collapse. For the four clusters we have studied, the outermost radius for which we have a mass density estimate should be included in the radial region where Cohn has found a constant slope. For M15 and 47 Tuc, the slopes for the regions beyond 2' agree well with the theoretical prediction (see Fig. 6), but both clusters exhibit a significant shoulder in the mass density profile inside of 2'. It will be interesting to see if such a significant shoulder appears during the Fokker-Planck simulations. For the two less concentrated clusters, NGC 3201 and NGC 362, the slopes are too shallow compared to the theoretical estimate in the outer parts of the clusters.

The phase-space distribution functions we derived are not consistent with King models. NGC 362 and NGC 3201 have significantly more stars which are tightly bound (low $E$) than King models predict (Fig. 8). For M15 and 47 Tuc there are systematic differences in the $f(E)$'s when compared to either the King models or the less concentrated clusters. Since M15 and 47 Tuc are considered possible core-collapse clusters, and NGC 362 and NGC 3201 are less concentrated, the systematic difference may be a signature of core-collapse. We have assumed isotropy for our determination of $f(E)$ which may not be valid during core-collapse (Cohn 1985).

We stress the importance of using non-parametric techniques to properly model one's data, even when it is sparsely sampled. To do otherwise may result in unknown biases and misinterpretation. The techniques described in this paper can be used for other dynamical systems, such as galactic nuclei and clusters of galaxies.

KG is grateful to Tad Pryor and D. Merritt for many discussions about the non-parametric techniques. We also thank Tad for generating MK models for M15 and 47 Tuc and for a thorough reading of the manuscript which greatly improved it. Pat Côté graciously supplied us with the velocity and surface brightness data for NGC 3201. KG wishes to acknowledge support from the National Need Fellowship of the U.S. Department of Education. Partial support of this research comes from grant AST-90-20685 from the National Science Foundation. PF thanks NSERC and Bell Labs for postdoctoral fellowships.

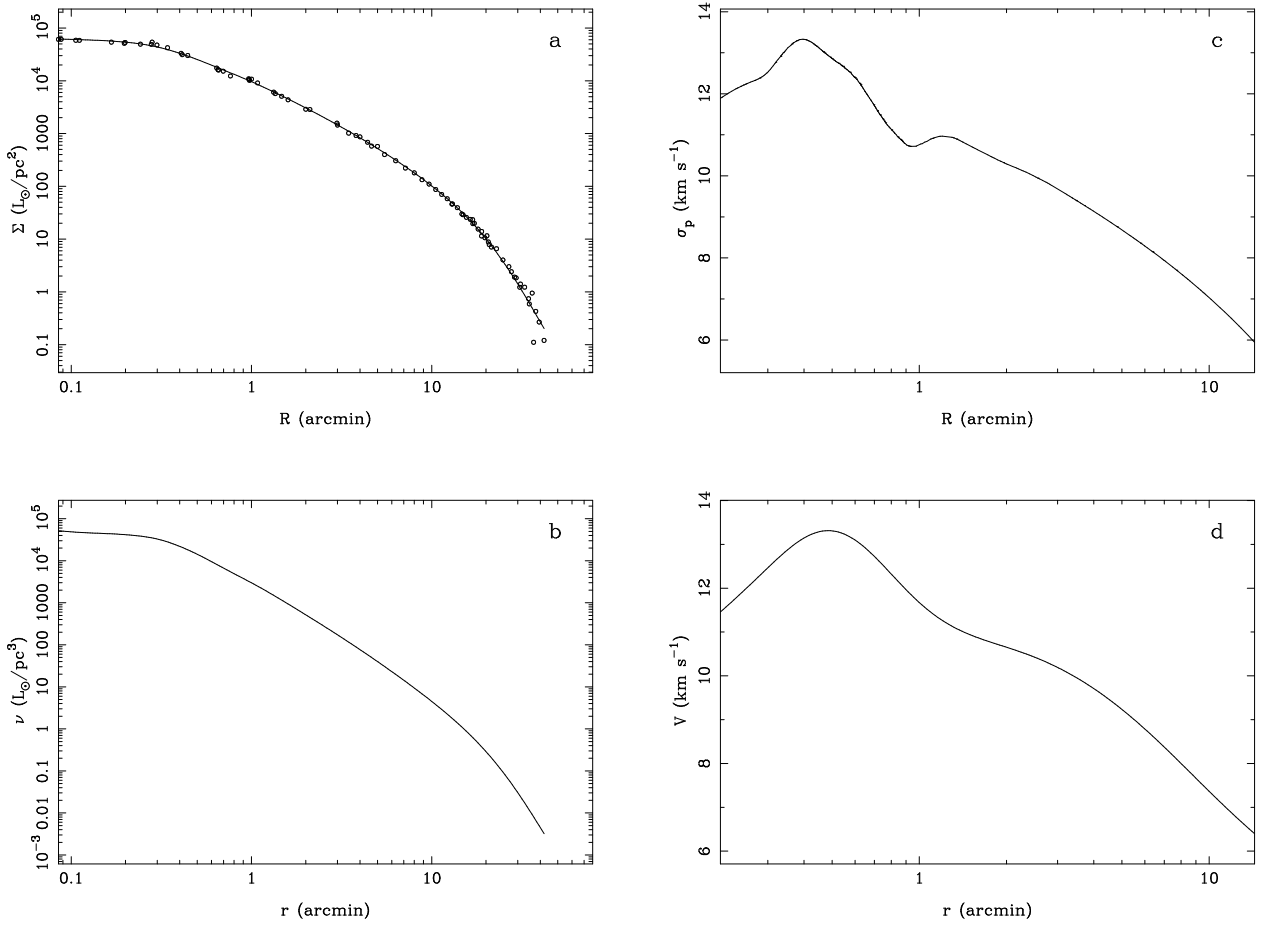

Fig. 1.— Projected and deprojected surface brightness and velocity dispersion profiles for 47 Tuc. The points in 1a are the surface brightness measurements from Meylan et al. (1988). The line in 1c is the projected velocity dispersion based on the radial velocities in Gebhardt et al. (1994b).

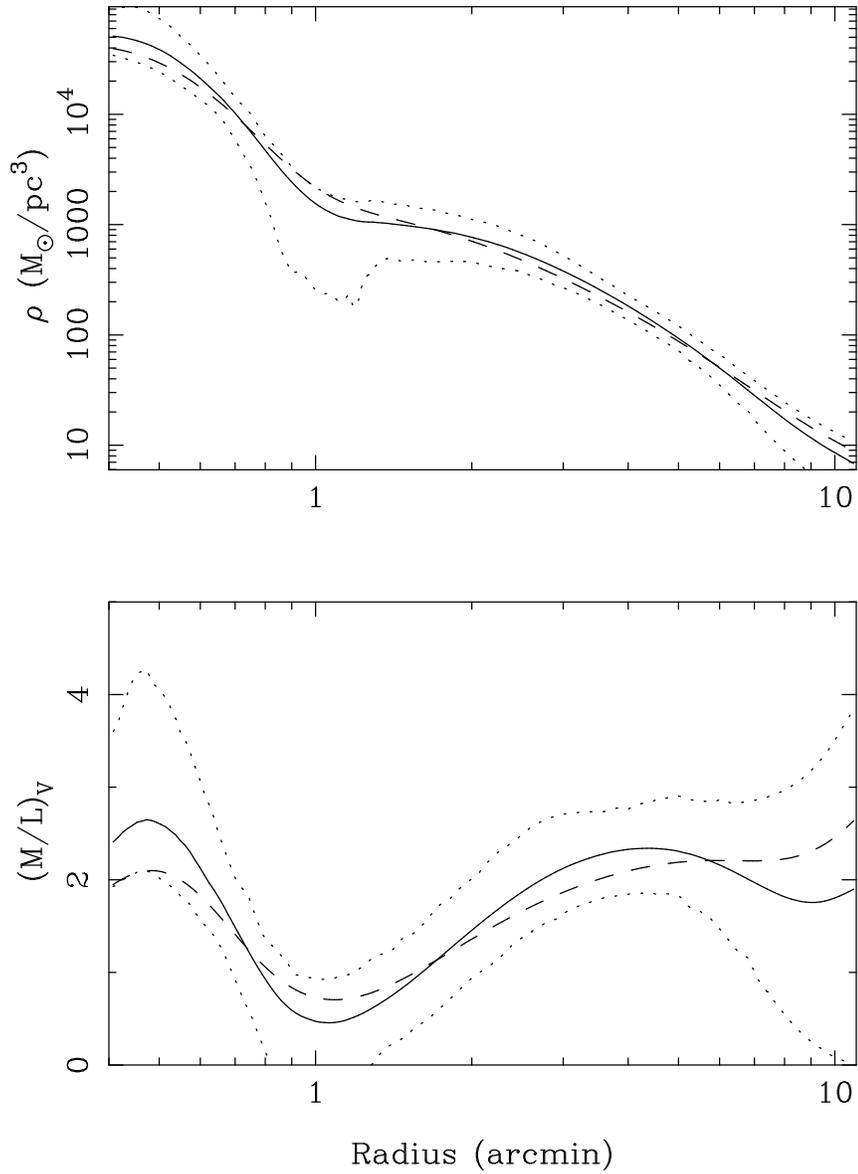

Fig. 2.— Mass density and M/L profiles for 47 Tuc. The solid lines are bias-corrected values, and the dashed lines are not bias corrected. The dotted lines are the bias-corrected 90% confidence bands. For 47 Tuc, $1' = 1.4$ pc.

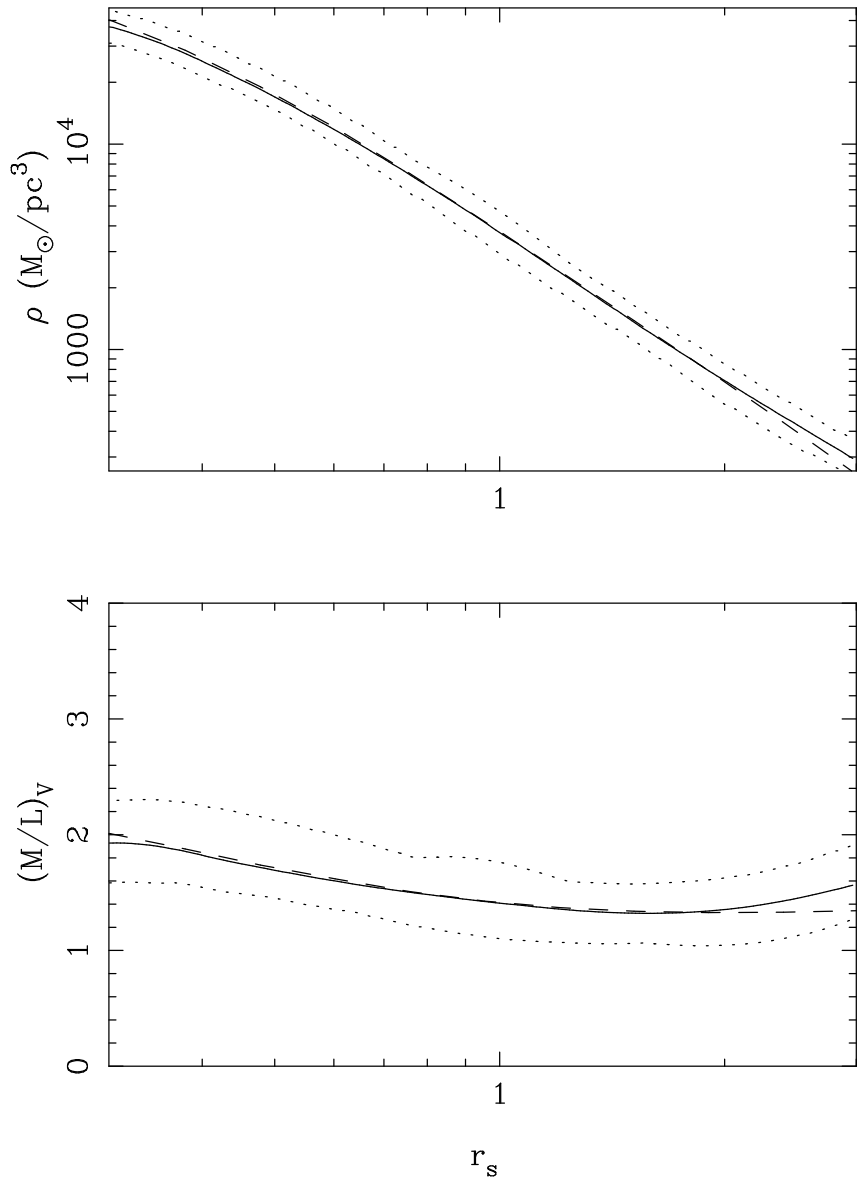

Fig. 3.— The average mass density and M/L profiles from 20 realizations of an isotropic multi-mass Michie-King model with mass function index $x = 0$. The solid lines are the averages and the dashed lines are the true profiles. The dotted lines are the 90% confidence bands for the 20 simulations.

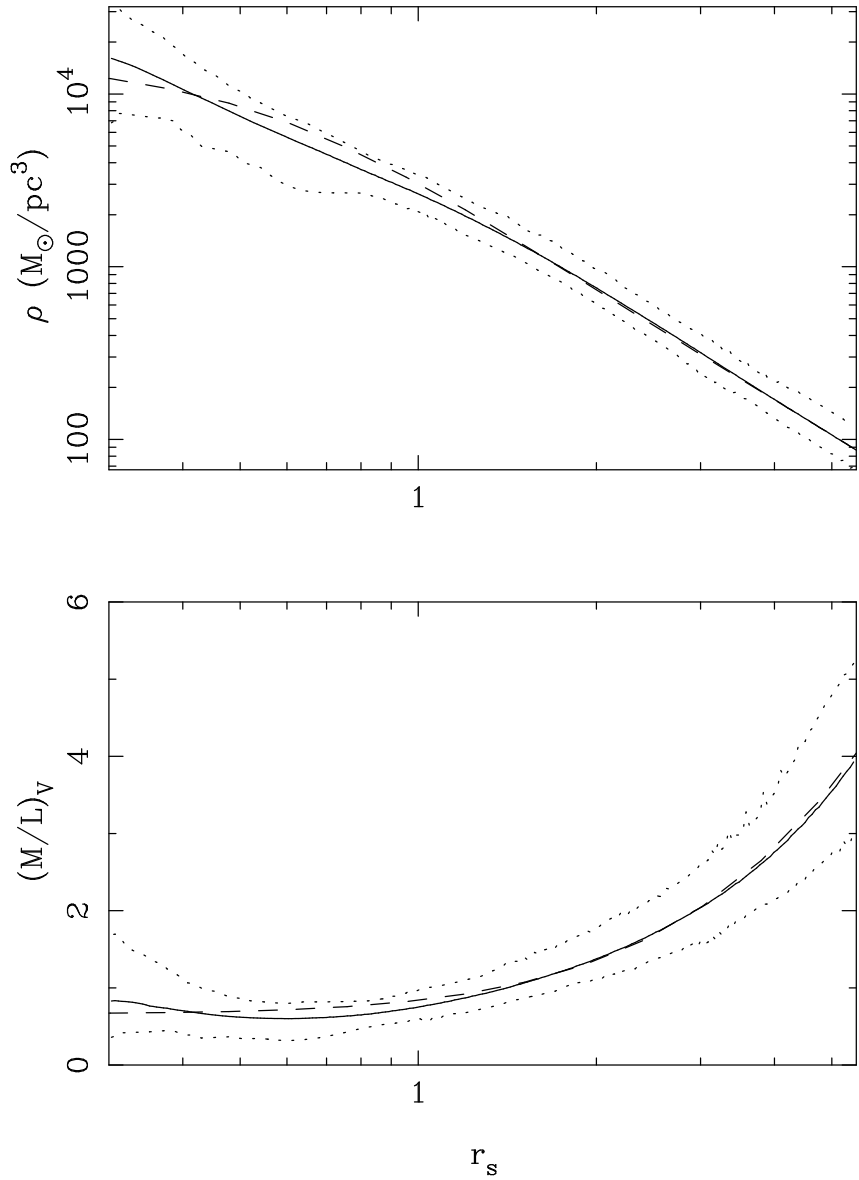

Fig. 4.— Mass density and M/L profiles for a single realization of an isotropic multi-mass Michie-King model with a mass function index $x = 2$. The solid lines are the inferred profiles and the dashed lines are the true profiles. The dotted lines are the 90% confidence bands.

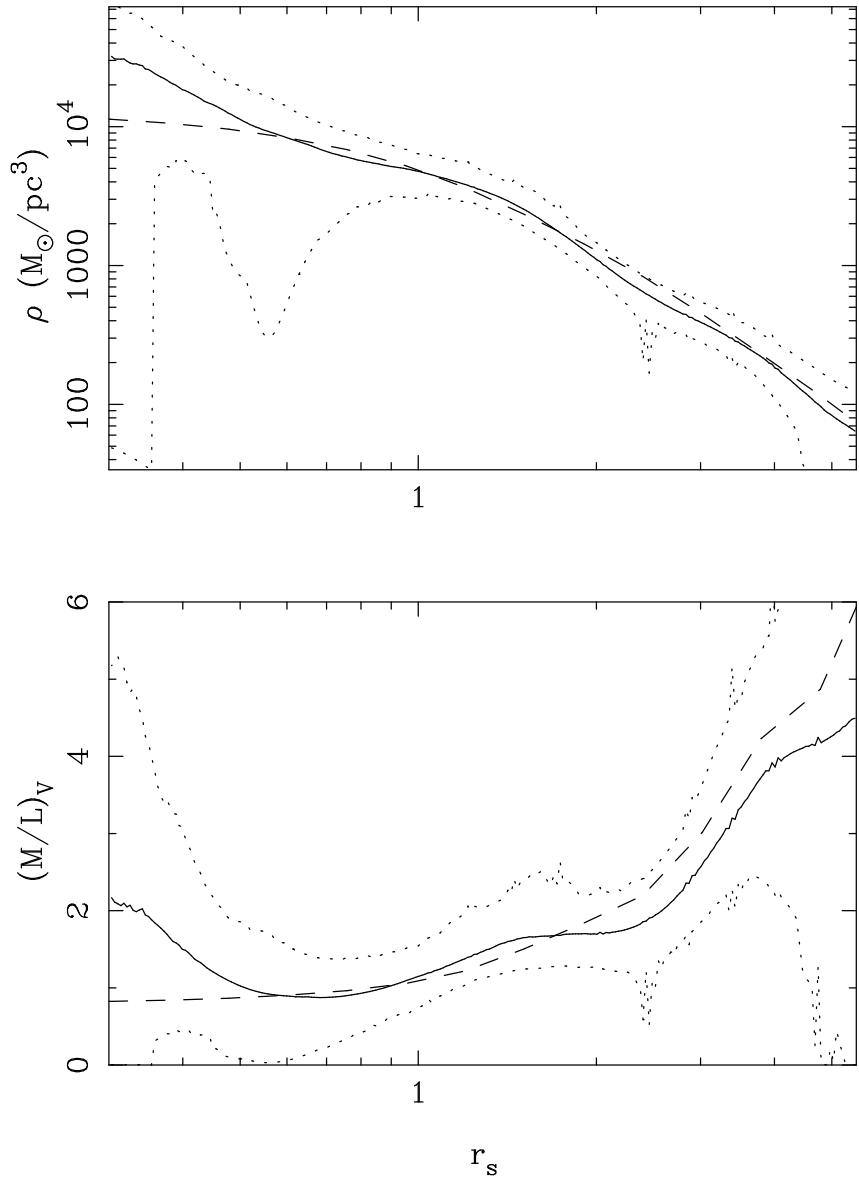

Fig. 5.— Mass density and M/L profiles for an anisotropic Michie-King model with $x = 2$ and anisotropy radius of $r_a = 3r_s$. The lines are the same as in Fig. 4.

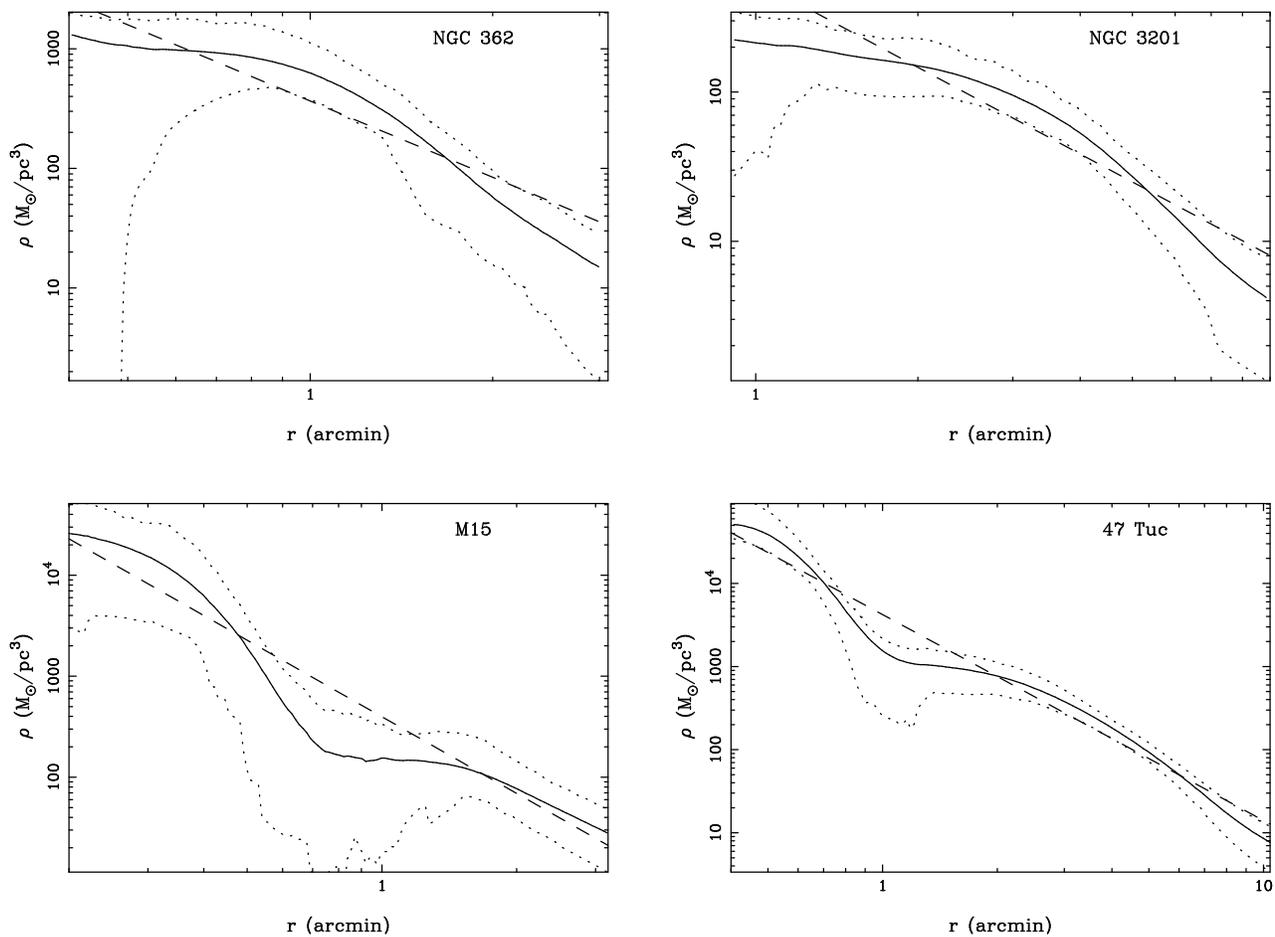

Fig. 6.— Non-parametric estimates of the mass density (solid lines) and the 90% confidence bands (dotted lines) for the four clusters. The dashed line is the theoretical slope for a core-collapse cluster. For NGC 362, $1' = 2.6$ pc, 1.4 pc for NGC 3201, 2.8 pc for M15, and 1.4 pc for 47 Tuc.

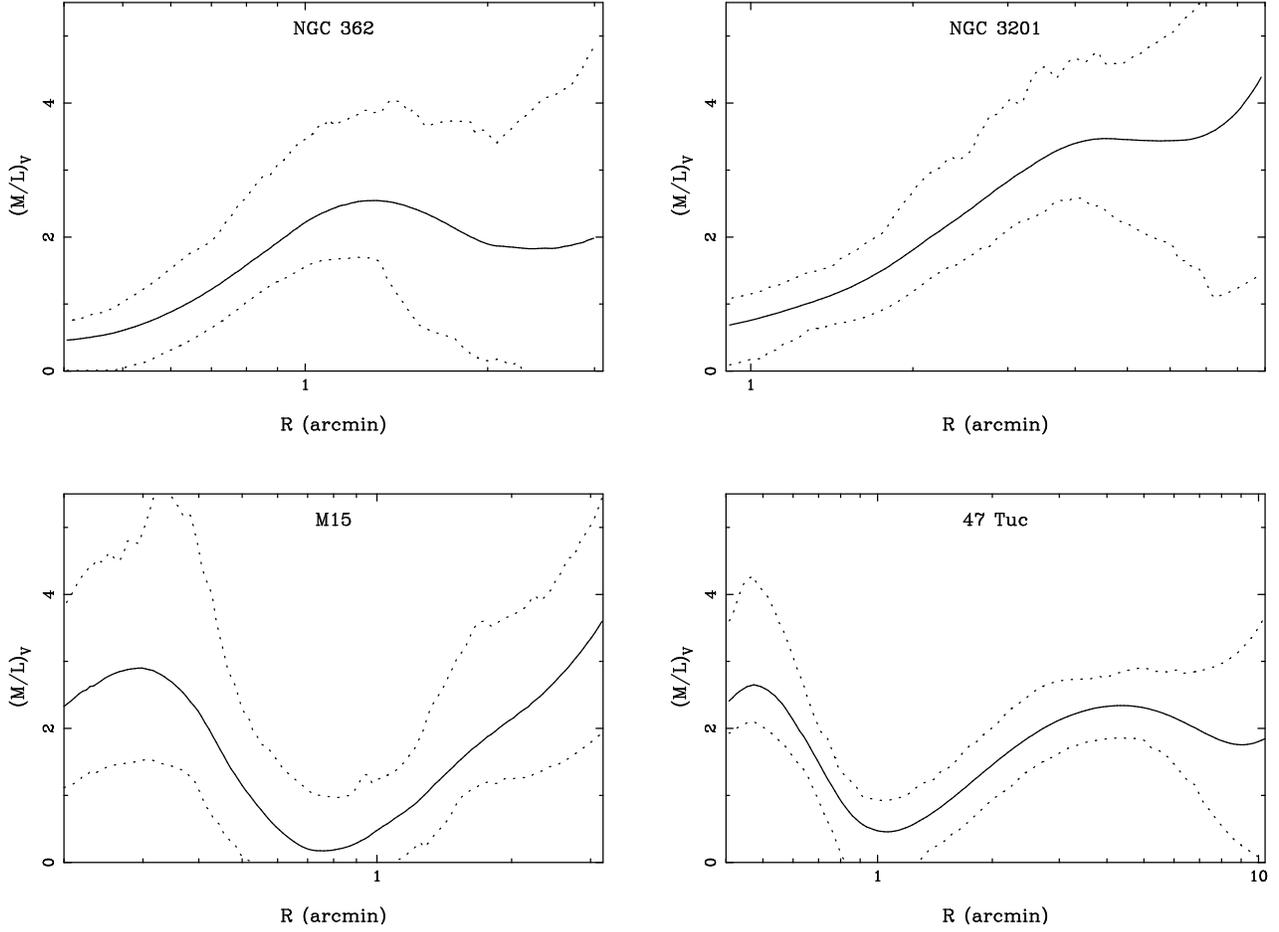

Fig. 7.— Non-parametric estimates of the M/L (solid lines) and the 90% confidence bands (dotted lines).

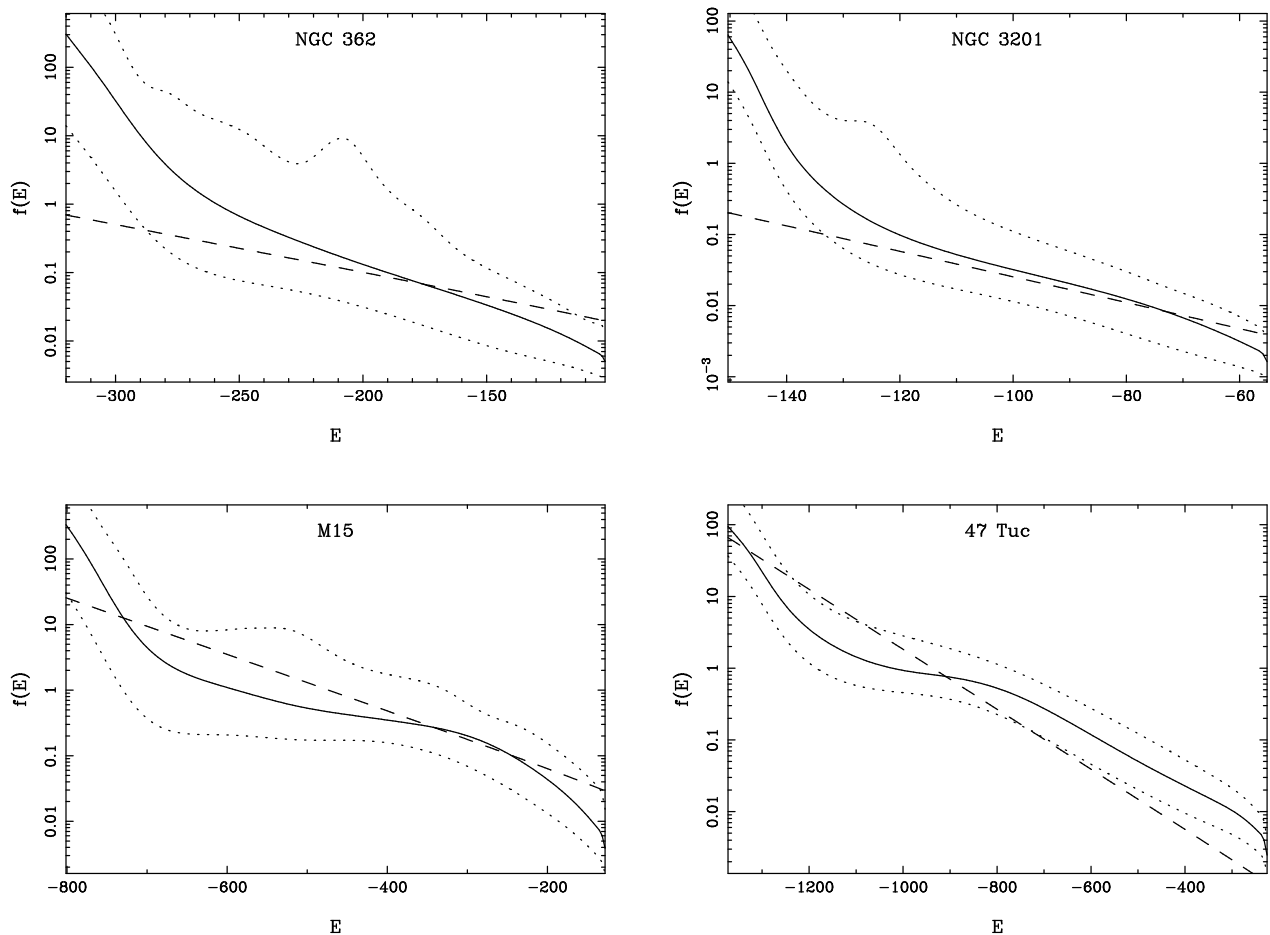

Fig. 8.— Phase space distribution functions for the tracer population. The solid line is the profile derived non-parametrically and the dashed line is the best fit isotropic Michie-King model from the literature. The dotted lines are the 90% confidence bands obtained via bootstrap. The units of energy are in (km s$^{-1}$)$^2$, and $f(E)$ is in units of number per (km s$^{-1}$)$^3$ per pc$^3$.

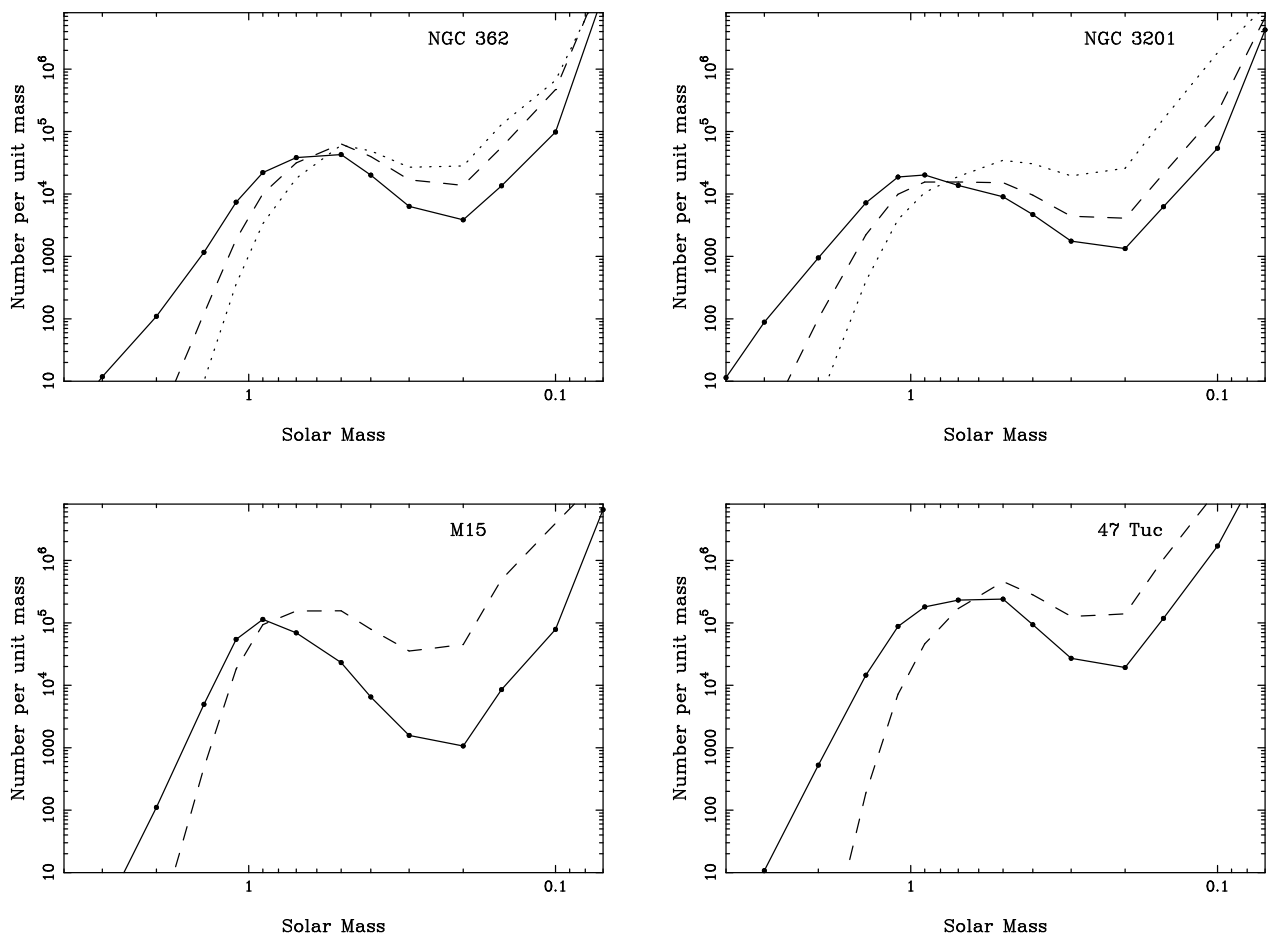

Fig. 9.— Mass functions. The different lines represent different radial regions: the solid line is the inner 25% mass of our total measured mass, the dashed line is the 25–50% mass, and the dotted line is the 50–70% mass.

TABLE 1. King Model Parameters

| Cluster | $\rho_0(M_\odot/\mathrm{pc}^3)$ | $r_s$ (pc) | $r_t$ (pc) | $M_t$ (M$_\odot$) | $\overline{m}_0$ |
|---------|---------------------------------|------------|------------|-------------------|------------------|
| NGC 362 | 3.3x10$^4$ | 0.50 | 30 | 3.3x10$^5$ | 0.9 |
| NGC 3201 | 6.9x10$^2$ | 2.04 | 45 | 1.9x10$^5$ | 1.0 |
| M15 | 4.2x10$^6$ | 0.05 | 24 | 4.8x10$^5$ | 1.1 |
| 47 Tuc | 3.3x10$^3$ | 0.70 | 70 | 1.5x10$^6$ | 0.8 |